\documentclass[aps,twocolumn,showpacs]{revtex4}
\usepackage{amsmath}
\usepackage{epsfig}
\usepackage{amssymb}
\begin{document}

\title{Geometric Engineering and Almost Mathieu Operator}

\newcommand*{\NJNU}{Department of Physics, Nanjing Normal University, Nanjing, Jiangsu 210097, China}\affiliation{\NJNU}

\author{Jing Zhou}\email{171001005@stu.njnu.edu.cn}\affiliation{\NJNU}
\author{Jialun Ping}\email{jlping@njnu.edu.cn (Corresponding author)}\affiliation{\NJNU}

\begin{abstract}
The type IIA string theory on a non-compact Calabi-Yau geometry known as the local $\mathbb{P}^{1} \times \mathbb{P}^{1}$ gives rise to five-dimensional N =1 supersymmetric SU(2) gauge theory
compactified on a circle, known as geometric engineering. So it is necessary to study the $\mathbb{P}^{1} \times \mathbb{P}^{1}$ in details. Since the spectrum of the local $\mathbb{P}^{1} \times \mathbb{P}^{1}$ can be written as $E=R^{2}\left(\mathrm{e}^{p}+\mathrm{e}^{-p}\right)+\mathrm{e}^{x}+\mathrm{e}^{-x}$, then by the result of almost Mathieu operator, we show that: (1) when $R^{2}<1$, the spectrum is absolutely continuous which meanings the medium is conductor. (2) when $1\le R^{2}<e^{\beta}$, the spectrum is singular continuous known as quantum Hall effect. (3) when $R^{2}>e^{\beta}$, the spectrum is almost surely pure point and exhibits Anderson localization. In other words, there are two phase transition points which one is $R^{2}=1$ and the other one is $R^{2}=e^{\beta}$.

\end{abstract}


\maketitle

\section{\label{sec:introduction}Introduction}
One of the most powerful consequences of our deep understanding of the string theory has been 
the appreciation of the fact that the gauge theory can be encoded geoemetrically in the structure 
of compactifications of type II string theory \cite{Kachru:1995wm,Bershadsky:1995sp}. 
Gauge groups arise through ADE singularities of geometry \cite{Klemm:1995tj}, however matter arises 
as loci of enhanced singularities. It not only leads to a unified description of gravitational and 
gauge theory, but it also leads directly to a deeper understanding of gauge dynamics, even in the 
limit of turning off gravitational effects \cite{Klemm:1996bj,Kachru:1995fv}. The basic idea is to 
geometrically engineer the gauge symmetry and matter content, and then study 
the corresponding theory using string techniques. As we know, the type IIA string theory on local 
Calabi-Yau geometry is known to give rise to five-dimensional $N =1$ supersymmetric $SU(2)$ gauge theory 
compactified on a circle, and as its mirror, the curve knows the non-perturbative information on 
this gauge theory. This curve reduces to the celebrated Seiberg-Witten \cite{Seiberg:1994rs} curve 
encoding the information on instantons in $N =2$ supersymmetric pure $SU(2)$ gauge theory. 
So it is natural to study the property of the local $\mathbb{P}^{1} \times \mathbb{P}^{1}$,
which is known as the non-compact Calabi-Yau geometry. 
In fact, we argue that there exist metal-insulator phase transition in the 
$\mathbb{P}^{1} \times \mathbb{P}^{1}$. It seem that there is no relation between string theory and
metal-insulator transition. However, the relation really exists.

In Ref.~\cite{Vafa:2015euh}, Vafa suggested a new model for fractional quantum Hall effect which 
connected it to the recent developments in string theory and led to new predictions for the principal 
series of fractional quantum Hall effect with filling fraction. And he found that the six dimensional 
$(2, 0)$ super-conformal field theory \cite{Witten:2009at} on the Riemann surface provided a realization 
of fractional quantum Hall effect in the $M$-theory. Moreover, he proposed that the $q$-deformed 
version of Chern-Simons theories were related to the anisotropic limit of fractional quantum Hall systems 
which split the zeroes of the Laughlin wave function. And he also suggested that the non-abelian 
topologically twisted Yang-Mills theory \cite{Vafa:1994tf} could realize as topological insulators. 
Now we can see the condense matter theory has been applied in string theory. So, we want to say that 
the string theory may have deep relation with the metal-insulator transition.

The almost Mathieu operator \cite{Jitomirskaya} is the discrete one dimensional Schr\"{o}dinger operator 
(acting on $\ell^{2}(Z)$),
\begin{equation}
H_{\alpha, \lambda, \theta} u(n)=u(n+1)+u(n-1)+2\lambda \cos (2 \pi \alpha n+\theta) u(n)
\end{equation}
where $\alpha, \lambda, \theta \in R$ are parameters, called the coupling, frequency, and phase. The name 
of the operator comes from the similarity to the Mathieu equation,
\begin{equation}
-y^{\prime \prime}(x)+\lambda \cos (x) y(x)=E y(x)
\end{equation}
$H_{\alpha, \lambda, \theta}$ is a tight-binding model for the Hamiltonian of an electron in a 
one- or two-dimensional lattice, subjecting to a perpendicular magnetic field. 
This model also describes a square lattice with anisotropic nearest neighbor coupling and isotropic 
next nearest neighbor coupling. The almost Mathieu operator plays an important role in the study of 
fundamental problems related to Bloch electrons in magnetic fields. In particular, it plays a major 
role in the Thouless-Kohmoto-Nightingale-den Nijs (TKNN) \cite{Thouless:1982zz} theory of the integer 
quantum Hall effect, where it gives rise to a rich set of possible integer Hall conductances. 
It is interesting to note that the TKNN theory which reflect the relevance of the almost Mathieu 
operator for describing electrons in magnetic fields has been verified experimentally \cite{Albrecht}.


In this work, we focus on the local $\mathbb{P}^{1} \times \mathbb{P}^{1}$ and almost Mathieu operator. 
And we hope that it may shed light on the issue of geometric engineering. The local $\mathbb{P}^{1} \times \mathbb{P}^{1}$ model is briefly introduced in Section II. And Section III is about almost Mathieu operators.
A brief summary is given in the last section.

\section{local $\mathbb{P}^{1} \times \mathbb{P}^{1}$}
Firstly, Let us consider the two-dimensional motion of electrons in the periodic potential and 
the magnetic field perpendicular to the two-dimensional plane. Then, in suitable limits, 
the Hamiltonian of the system is described by,
\begin{equation}
H=\mathrm{e}^{\mathrm{i} x}+\mathrm{e}^{-\mathrm{i} x}+\mathrm{e}^{\mathrm{i} p}+\mathrm{e}^{-\mathrm{i} p},
\end{equation}
The spectrum is an intricate pattern which study by Hofstadter, known as Hofstadter¡¯s butterfly. 
And this system was later used as a model system where the topological numbers determine the quantum 
Hall conductance. Actually, in high energy physics, a similar equation has been widely
studied \cite{Hatsuda:2016mdw},
\begin{equation}
H=\mathrm{e}^{x}+\mathrm{e}^{-x}+\mathrm{e}^{p}+\mathrm{e}^{-p}
\end{equation}
When $x$ and $p$ are restricted to be purely imaginary, then this equation reduces to Hofstadter¡¯s 
Hamiltonian. In fact, this equation arises when geometric engineering is applied to a non-compact 
Calabi-Yau geometry known as the local $\mathbb{P}^{1} \times \mathbb{P}^{1}$ geometry. 
The Kahler structure of the original Calabi-Yau is mapped to the complex structure of the 
mirror Calabi-Yau, and vice versa. The mirror Calabi-Yau to local $\mathbb{P}^{1} \times \mathbb{P}^{1}$ 
is described by the following equation \cite{Huang:2014nwa,Huang:2014eha},
\begin{equation}
\mathrm{e}^{x}+z_{1} \mathrm{e}^{-x}+\mathrm{e}^{p}+z_{2} \mathrm{e}^{-p}=1
\end{equation}
where $z_{1}$ and $z_{2}$ are the complex moduli of the mirror Calabi-Yau space. The mirror curve has 
many information to construct the all-genus free energy of the topological string 
theory \cite{Marino:2006hs,Marino:2009jd,Kozcaz:2010af}. For our purpose, we have,
\begin{equation}
x \rightarrow x-\log E, \quad p \rightarrow p+2 \log R-\log E
\end{equation}
and to set,
\begin{equation}
z_{1}=\frac{1}{E^{2}}, \quad z_{2}=\frac{R^{4}}{E^{2}}
\end{equation}
Then the Eq.~(5) can be rewritten as,
\begin{equation}
R^{2}\left(\mathrm{e}^{p}+\mathrm{e}^{-p}\right)+\mathrm{e}^{x}+\mathrm{e}^{-x}=E
\end{equation}
So, we can write the local $\mathbb{P}^{1} \times \mathbb{P}^{1}$ as Eq.~(8)

\section{Almost Mathieu Operator}
As we know, the origin almost Mathieu operator is,
\begin{equation}
H_{\alpha, \lambda, \theta} u(n)=u(n+1)+u(n-1)+2\lambda \cos (2 \pi \alpha n+\theta) u(n)
\end{equation}
For our purpose, we can rewrite it as,
\begin{equation}
R^{2}\left(\mathrm{e}^{p}+\mathrm{e}^{-p}\right)+\mathrm{e}^{x}+\mathrm{e}^{-x}=E
\end{equation}
where $R^2=\lambda$. So, it is the same with local $\mathbb{P}^{1} \times \mathbb{P}^{1}$ model. 
And one can find that when $R=1$, this equation reduces to Hofstadter's Hamiltonian.

In fact, the spectrum of $E$ is widely researched in mathematics. And one can find that the spectrum 
is depending on the coupling parameter $\lambda$. Actually, It is known that ,

(1) For $0<\lambda <1$, $E$ has surely purely absolutely continuous spectrum \cite{Avila}. 
In physics, we can regard the particle "travels freely" in the medium. In this case, the medium is conductor.

(2) For $1\leq\lambda<e^{\beta}$, $E$ has almost surely purely singular continuous spectrum. It is 
the quantum Hall effect.

(3) For $\lambda>e^{\beta}$, $E$ has almost surely pure point spectrum \cite{youjiangong} and exhibits 
Anderson localization. Or in other words, the medium is insulator

Anderson localization \cite{Anderson:1958vr,Pixley:2015xla} can be understood as an interference phenomenon. 
In this case, the materials can undergo a phase transition from conductor to insulator. In the original
tight-binding model formulated by Anderson, the electrons are able to tunnel between neighbouring lattice 
sites. However, at high enough disorder in the lattice, the quantum amplitudes associated with tunnelling 
paths cancel each other, resulting in a localized wave function,
and in the way, the scattered wavelets interfere destructively in the forward direction,
causing the wave to decay exponentially.

Now, one may find that the property of local $\mathbb{P}^{1} \times \mathbb{P}^{1}$ highly depends on 
coupling parameter $\lambda$. When the coupling parameter is small, the medium is in order, then the 
particle travels freely. And when the coupling parameter is large, the medium behaves as insulator, 
then the particle do not travel.

\section{Summary}
We introduce the local $\mathbb{P}^{1} \times \mathbb{P}^{1}$ model and almost Mathieu operators. 
Then by the result of almost Mathieu operator, we show that there exist metal-instructor transition 
in the local $\mathbb{P}^{1} \times \mathbb{P}^{1}$ model. And there are two phase transition points 
which one is $R^{2}=1$ and another is $R^{2}=e^{\beta}$ in the local 
$\mathbb{P}^{1} \times \mathbb{P}^{1}$ model.

It is interesting that Vafa suggests the potential experimental realization of topologically 
twisted non-abelian Yang-Mills theories as effective descriptions in the bulk of topological 
insulators in lab by a suitable model. And Vafa and his coauthor are currently studying potential 
applications of this to condensed matter systems.

So, in our case, we argue that there may exist metal-insulator transition in the geometric engineering. 
And we hope that our comment will provide a new view for the studying of string theory

\section*{Acknowledgment}
Part of the work was done when Jing Zhou visited the Yau Mathematical Science Center.
Jing Zhou thanks for discussing with An Qi. This work is partly supported by the National Science Foundation of China under Contract Nos. 11775118, 11535005.

\end{document}